\documentclass[12pt]{article}

\usepackage{amsmath,amsthm,amssymb,euscript,amscd}
\usepackage[english]{babel} % espanol, 
\usepackage[utf8]{inputenc} % acentos sin codigo
\usepackage{graphicx}
\usepackage{url}
\usepackage{hyperref}
\usepackage{bm}
\usepackage{rotating}
\usepackage{fancyhdr}
\title{A Simplified Formulation for the Backward/Forward Sweep Power Flow Method}
\author{Paulo~M.~De~Oliveira-De Jesus\thanks{ \scriptsize
Electrical \& Electronic Engineering Department, School of Engineering, Universidad de los Andes, Colombia, website: \url{https://power.uniandes.edu.co/pdeoliveira}, E-mail: \url{pm.deoliveiradejes@uniandes.edu.co}} }

\begin{document}

\maketitle

\begin{abstract}
    This paper describes a simplified formulation of the
    Backward/Forward (BW/FW) Sweep Power Flow applied to radial
    distribution systems with distributed generation under positive sequence modelling. Proposed formulation was applied in an illustrative test system.

\end{abstract}
\section{Introduction}

Several Backward/Forward (BW/FW) sweep algorithms have
been discussed in literature. In 1967, Berg presented a paper which
can be considered as the source for the all variants of BW/FW sweep
methods \cite{berg}. Later, a similar approach was presented in
\cite{kersting} based on ladder network theory. The BW/FW Sweep
algorithms use the Kirchhoff laws. Different formulations can be
found
\cite{shirmo,gosh,fuku,thuka,jova,raji,aravi2,zhu,liu,afsari,teng,chang,prasad}.
BW/FW sweep methods typically present a slow convergence rate but
computationally efficient at each iteration. Using these methods,
power flow solution for a distribution network can be obtained
without solving any set of simultaneous equations. In this work, the
standard BW/FW sweep power flow is reformulated in convenient form.
An illustrative four-bus example is solved.

\section{The Method}

 The input data of this algorithm is given by
node-branch oriented data used by most utilities. Basic data
required is: active and reactive powers, nomenclature for sending
and receiving nodes, and positive sequence impedance model for all
branches.

\begin{figure}[!t]
\centering
\includegraphics[width=4.6in]{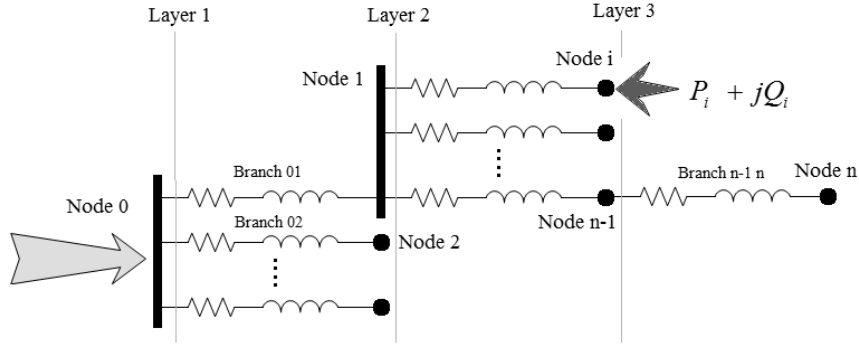}
% where an .eps filename suffix will be assumed under latex,
% and a .pdf suffix will be assumed for pdflatex; or what has been declared
% via \DeclareGraphicsExtensions.
\caption{Branch and node numbering of a radial distribution network}
\label{layers}
\end{figure}

In the following, the standard BW/FW sweep power flow method is
 written in matrix notation using complex variables. Branch impedances are
stated as a vector $\textbf{Z}$ corresponding to a distribution line
model containing a series positive sequence impedance for line or
transformer. Shunt impedances are not considered in this first
approach. Fig. \ref{layers} shows a radial distribution network with
$n+1$ nodes, and $n$ branches and a single voltage source at the
root node 0. Branches are organized according to an appropriate
numbering scheme (list), which details are provided in
\cite{shirmo}.

%\textbf{R}+j\textbf{X}=

\begin{equation}\label{1}
    \textbf{Z}=\left [\begin{array}{ccccc}
                                          \overline{Z}_{01} & ... & \overline{Z}_{ij} & ... & \overline{Z}_{mn} \\
                                        \end{array}
                                      \right]
\end{equation}
where,
\begin{equation}\label{2}
    \overline{Z}_{ij}=R_{ij}+jX_{ij} \quad i,j=1,...,n \quad i \neq j
\end{equation}

Bus data is given by

\begin{equation}\label{3}
      \mathbf{S}=\left[
        \begin{array}{c}
          \overline{S}_{1} \\
          \vdots \\
          \overline{S}_{i} \\
          \vdots \\
          \overline{S}_{n} \\
        \end{array}
      \right]= \left[
        \begin{array}{c}
          {P}_{1}+j{Q}_{1} \\
          \vdots \\
          {P}_{i}+j{Q}_{i} \\
          \vdots \\
          {P}_{n}+j{Q}_{n} \\
        \end{array}
      \right]
\end{equation}

where net nodal active and reactive powers are given by generated
and demanded powers:
\begin{align}
P_{i}&={P}_{Gi}-{P}_{Di}\\
Q_{i}&={Q}_{Gi}-{Q}_{Di}
\end{align}

The numbering of branches in one layer begins only after all the
branches in the previous layer have been numbered. Considering that
initial voltages are known: voltage at substation is set
$\overline{V}_{0}=Vref $ and an initial voltage vector is given by:

\begin{equation}\label{4}
    \mathbf{V}^0=\left [\begin{array}{ccccc}
                                          \overline{V}^0_{1} & ... & \overline{V}^0_{i} & ... & \overline{V}^0_{n} \\
                                        \end{array}
                                      \right]
\end{equation}
The state of the system is reached solving two steps iteratively.
\subsection{Step 1 - Backward Sweep}
For each iteration $k$, branch currents are aggregated from loads to
origin:
\begin{equation}\label{5}
    \mathbf{J^k}=-\mathbf{T} \cdot \mathbf{I^k}
\end{equation}

The relationship between nodal currents $\mathbf{I^k}$ and branch
currents $\mathbf{J^k}$ is set through an upper triangular matrix
$\mathbf{T}$ accomplishing the Kirchhoff Current Laws (KCL). Each
element $\overline{I}^k_{i}$ of $\mathbf{I^k}$ associated to node
$i$ is calculated as function of injected powers $\overline{S}_{i}$
and its voltage profile $\overline{V}^k_{i}$ as shown below:

\begin{equation}\label{6a}
    \overline{I}^k_{i}=\frac{\overline{S}^\ast_{i}}{\overline{V}^{k\ast}_{i}}
    \quad i=1,...,n
\end{equation}

\subsection{Step 2 - Forward Sweep}

Nodal voltage vector $\mathbf{V}$ is updated from the origin to
loads according the Kirchhoff Voltage Laws (KVL), using previously
calculated branch currents vector $\mathbf{J}$,  branch impedances
vector $\mathbf{Z}$:

\begin{equation}\label{6}
   \mathbf{V}^{k+1}=\mathbf{V}_0-\mathbf{T}^T \cdot \mathbf{D_Z} \cdot \mathbf{J}^k
\end{equation}
where $\mathbf{V}_0$ is a $n$-elements vector with all entries set
at voltage at origin (swing node) $\overline{V}_0$
   and branch impedances $\mathbf{D_Z}$ is the diagonal matrix of vector $\mathbf{Z}$:

Using Eq. \ref{5}

\begin{equation}\label{6}
   \mathbf{V}^{k+1}=\mathbf{V}_0+\mathbf{T}^T \cdot \mathbf{D_Z} \cdot \mathbf{T} \cdot \mathbf{I^k}
\end{equation}

Updated voltages can be updated using only one equation:

\begin{equation}\label{6}
   \mathbf{V}^{k+1}=\mathbf{V}_0+\mathbf{TRX} \cdot \mathbf{I^k}
\end{equation}

where $\mathbf{TRX}=\mathbf{T}^T \cdot \mathbf{D_Z} \cdot
\mathbf{T}$

\subsection{Convergence}

Updated voltages are compared with previous voltages in order to
perform convergence check in.

\begin{equation}\label{7}
\varepsilon \leq |\overline{V}^{k+1}_i-\overline{V}^{k}_i| \quad
i=1,...,n
\end{equation}

\section{Illustrative Example: Simply 4-node Network}

\begin{figure}[!h]
\centering
\includegraphics[width=2.2in]{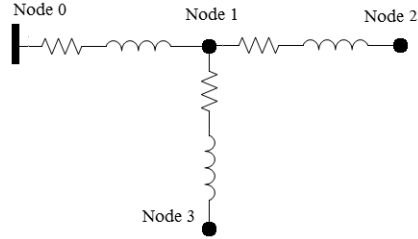}
% where an .eps filename suffix will be assumed under latex,
% and a .pdf suffix will be assumed for pdflatex; or what has been declared
% via \DeclareGraphicsExtensions.
\caption{4-Node Network Topology} \label{3bus}
\end{figure}

To illustrate the proposed methodology, it is used the 4-node
example shown in Fig. \ref{3bus}. Length of all sections is 1 mile.
Load demand at nodes 2 and 3 are 2MW with $\cos\varphi=1.0$.

Using the following bases $S_{B}=$10MW and $V_{B}$=12.47kV, data and
results are given in per unit. Loads are 0.2 in nodes 2 and 3.
Reference voltage at node 0 is $\overline{V}_0 = 1+j0$ and initial
voltages are set $\mathbf{V}^0=\left[\begin{array}{ccc}
1+0j & 1+0j & 1+0j \\
\end{array}
\right]$.

Branches are represented by:

\begin{equation}\label{30}\nonumber
    \mathbf{Z}=\mathbf{R}+j\mathbf{X}=\left[
             \begin{array}{c}
               .0296 \\
               .0296 \\
               .0296 \\
             \end{array}
           \right] +j\left[
             \begin{array}{c}
               .0683 \\
               .0683 \\
               .0683 \\
             \end{array}
           \right]
\end{equation}

Network topology is represented through a 3x3 upper triangular
matrix $\mathbf{T}$.

\begin{equation}\label{31}\nonumber
    \mathbf{T}=\left[
                 \begin{array}{ccc}
                   1 & 1 & 1 \\
                   0 & 1 & 0 \\
                   0 & 0 & 1 \\
                 \end{array}
               \right]
\end{equation}

Then, $\mathbf{D_Z}$ is:

\begin{equation}\label{31}\nonumber
\mathbf{D_Z}=\left[
                 \begin{array}{ccc}
                   .0296+j.0683 & .0 & 0 \\
                   0 & .0296+j.0683 & 0 \\
                   0 & 0& .0296+j.0683 \\
                 \end{array}
               \right]
\end{equation}

Solution reached at iteration 3 for $\varepsilon=10^{-4}$ and
displayed in Table \ref{unb1}. Results are presented in per unit and
degrees.

\begin{table}[!h]
\caption{4 Node State of the System - balanced Approach}
\begin{center}
\begin{tabular}{c c c c c c c c}
$V_0$ & $\theta_0$ & $V_1$ & $\theta_1$ & $V_2$ & $\theta_2$ & $V_3$ & $\theta_3$ \\
\hline 1.000 & 0.00 & 0.987 & -1.59 & 0.981 & -2.40 & 0.981 & -2.40
\\ \hline
\end{tabular}
\end{center}
\label{unb1}
\end{table}

\section{Conclusion}
    This paper describes a convenient formulation of the
    Backward/Forward (BW/FW) Sweep Power Flow applied to radial
    distribution systems with distributed generation. Proposed formulation was applied in an illustrative test system.

\section{Nomencalture}

\emph{List of Symbols}
\begin{tabbing}
 texto  \= 1 \kill
  % \> for next tab, \\ for new line...

  $\mathbf{D_{Z}}$ \> Diagonal matrix
of branch impedance vector
$\mathbf{Z}$ \\
 $\mathbf{{R}}$  \> Diagonal matrix of branch resistance vector
$\Re e {\mathbf{Z}}$ \\
  $\mathbf{{X}}$ \> Diagonal matrix of branch reactance vector
$Im {\mathbf{Z}}$ \\
$\varepsilon$ \> Convergence criteria \\
$\mathbf{I}$ \> Current vector \\
$\mathbf{J}$ \> Branch Current vector $\mathbf{J}$ \\
$n$ \> Number of nodes, excluding origin \\
$\mathbf{P}$ \> Active Power Injected vector \\
$\mathbf{Q}$ \> Reactive Power Injected vector \\
$P_{j}$ \> Active Power Injected at node $j$ \\
$Q_{j}$ \> Reactive Power Injected at node $j$ \\
$P_{Dj}$ \> Active Power Demanded at node $j$ \\
$Q_{Dj}$ \> Reactive Power Demanded at node $j$ \\
$P_{Gj}$ \> Active Power Generated at node $j$ \\
$Q_{Gj}$ \> Active Power Generated at node $j$ \\
$R_{ij}$ \> Resistance between node $i$ and node $j$\\
$S_{Dj}$ \> Apparent Power Demanded at node $j$ \\
$S_{Gj}$ \> Apparent Power Generated at node $j$ \\
$\mathbf{T}$ \> Triangular matrix\\
$\mathbf{V}$ \> Voltage vector \\
$X_{ij}$ \> Reactance between node $i$ and node $j$\\
$\mathbf{Z}$ \> Branch Impedance vector $\mathbf{Z}$ \\
$\overline{Z}^{ij}$ \> Branch Impedance between node $i$ and node $j$\\
$\mathbf{Z}^{ij}$ \>  Impedance matrix between node $i$ and node $j$\\
\end{tabbing}

\emph{Operators}
\begin{tabbing}
% texto  \= 1 \kill
%  % \> for next tab, \\ for new line...%
%
$T\quad\quad$ \= Transpose Matrix \\
$D$ \> Diagonal Matrix \\
$*$ \> Conjugate of a complex number
\end{tabbing}

\emph{Sub-Indexes}
\begin{tabbing}
% texto  \= 1 \kill
%  % \> for next tab, \\ for new line...
$i\quad\quad$ \= Associated to node  $i$\\
$j$ \> Associated to node  $j$ \\
$k$ \> Associated to iteration  $k$ \\
\end{tabbing}

\end{document}